\documentclass{cupconf}
\usepackage{epsfig}

\title[Eruptive Mass Loss and the First Stars]{Eruptive Mass Loss in
  Very Massive Stars and Population III Stars}

\author[N.\ Smith]{N\ls A\ls T\ls H\ls A\ls N\ns S\ls M\ls I\ls T\ls
  H}

\affiliation{Center for Astrophysics and Space Astronomy, University
of Colorado, 389 UCB, Boulder, CO 80309, USA;
nathans@colorado.edu}

\begin{document}
\maketitle

\begin{abstract}

I discuss the role played by short-duration eruptive mass loss in the
evolution of very massive stars.  Giant eruptions of Luminous Blue
Variables (LBVs) like the 19th century event of $\eta$ Carinae can
remove large quantities of mass almost instantaneously, making them
significant in stellar evolution.  They can potentially remove much
more mass from the star than line-driven winds, especially if stellar
winds are highly clumped such that previous estimates of O star
mass-loss rates need to be revised downward.  When seen in other
galaxies as ``supernova impostors'', these LBV eruptions typically
last for less than a decade, and they can remove of order 10
M$_{\odot}$ as indicated by massive nebulae around LBVs.  Such extreme
mass-loss rates cannot be driven by radiation pressure on spectral
lines, because the lines will completely saturate during the events.
Instead, these outbursts must either be continuum-driven
super-Eddington winds or outright hydrodynamic explosions, both of
which are insensitive to metallicity.  As such, this eruptive mode of
mass loss could also play a pivotal role in the evolution and ultimate
fate of massive metal-poor stars in the early universe.  If they occur
in these Population III stars, such eruptions would also profoundly
affect the chemical yield and types of remnants from early supernovae
and hypernovae thought to be the origin of long gamma ray bursts.

\end{abstract}

\section{Introduction}

Mass loss is a critical factor in the evolution of a massive star.  In
addition to the direct reduction of a star's mass, it profoundly
affects the size of its convective core, its core temperature, its
angular momentum evolution, its luminosity as a function of time, and
hence its evolutionary track on the HR diagram and its main-sequence
(MS) lifetime (e.g., Chiosi \& Maeder 1986).  Wolf-Rayet (WR) stars
are thought to be the descendants of massive stars as a consequence of
mass loss in the preceding H-burning phases, during which the star
sheds its H envelope (Abbott \& Conti 1987; Crowther 2006).  While the
maximum initial mass of stars is thought to be $\sim$150 M$_{\odot}$
(Figer 2005; Kroupa 2005), WR stars do not have masses much in excess
of 20 M$_{\odot}$ (Crowther 2006).\footnote{By ``WR stars'' we mean
H-deficient WR stars (core-He burning phases or later), and not the
luminous H-rich WNL stars (Crowther et al.\ 1995), which are probably
still core-H burning.}  Thus, very massive stars have the immense
burden of removing 30--130 M$_{\odot}$ during their lifetime before
the WR phase, unless they explode first.  Stellar evolution
calculations prescribe $\dot{M}$($t$) based on semiempirical values,
so we need to know when this mass loss occurs.

The main question I wish to address here is whether the majority of
mass lost during the lifetime of the most massive stars occurs
primarily via steady line-driven stellar winds, or instead through
violent, short-duration eruptions or explosions.  The two extremes are
shown graphically in Figure 1.  This question is critical for
understanding how mass loss scales with metallicity, back to the time
of the massive stars in the early Universe.  In this contribution I
would like to draw attention to the specific role of LBV eruptions,
advocating for their importance.  The essential points of the argument
are the following:

\begin{itemize}

\item Recent studies of hot star winds indicate that mass-loss rates
  on the MS are probably much lower than previously thought due to the
  effects of clumping in the wind.  If so, then for the most massive
  stars, the revised mass-loss rates are inadequate to reduce the
  star's mass enough to reach the WR phase, where the H-rich envelope
  has been removed and the He-rich core material is exposed. Something
  other than the O-star's line-driven wind must account for this mass
  loss.

\item Observations of nebulae around luminous blue variables (LBVs)
  and LBV candidates have revealed very high ejecta masses -- of order
  10 M$_{\odot}$. In some objects, there is evidence for multiple
  shell ejections on timescales of 10$^3$ years.  Cumulatively, these
  sequential eruptions could, in principle, remove a large fraction of
  the total mass of the star.  Thus, a few short-duration outbursts
  like the 19th century eruption of $\eta$ Car could dominate mass
  lost during the lives of the most massive stars, and would be
  critical for the envelope shedding needed to form WR stars at any
  metallicity.

\item The extreme mass-loss rates of these LBV bursts imply that line
  opacity is too saturated to drive them, so they must instead be
  either continuum-driven super-Eddington winds (see Owocki et al.\
  2004) or outright hydrodynamic explosions.  Unlike steady winds
  driven by lines, the driving in these eruptions may be largely
  independent of metallicity, and might play a role in the mass loss
  of massive metal-poor stars (Population III stars).

\end{itemize}

These points have already been explained and justified in more detail
by Smith \& Owocki (2006), where more complete references can be
found.  That original paper was deliberately provocative, and some
caveats had to be left out for the sake of brevity.  Rather than
repeat that discussion, I will briefly elaborate on a few of these
issues, and will spend most of the time discussing alternatives and
further implications.

\begin{figure}\begin{center}
\epsfig{file=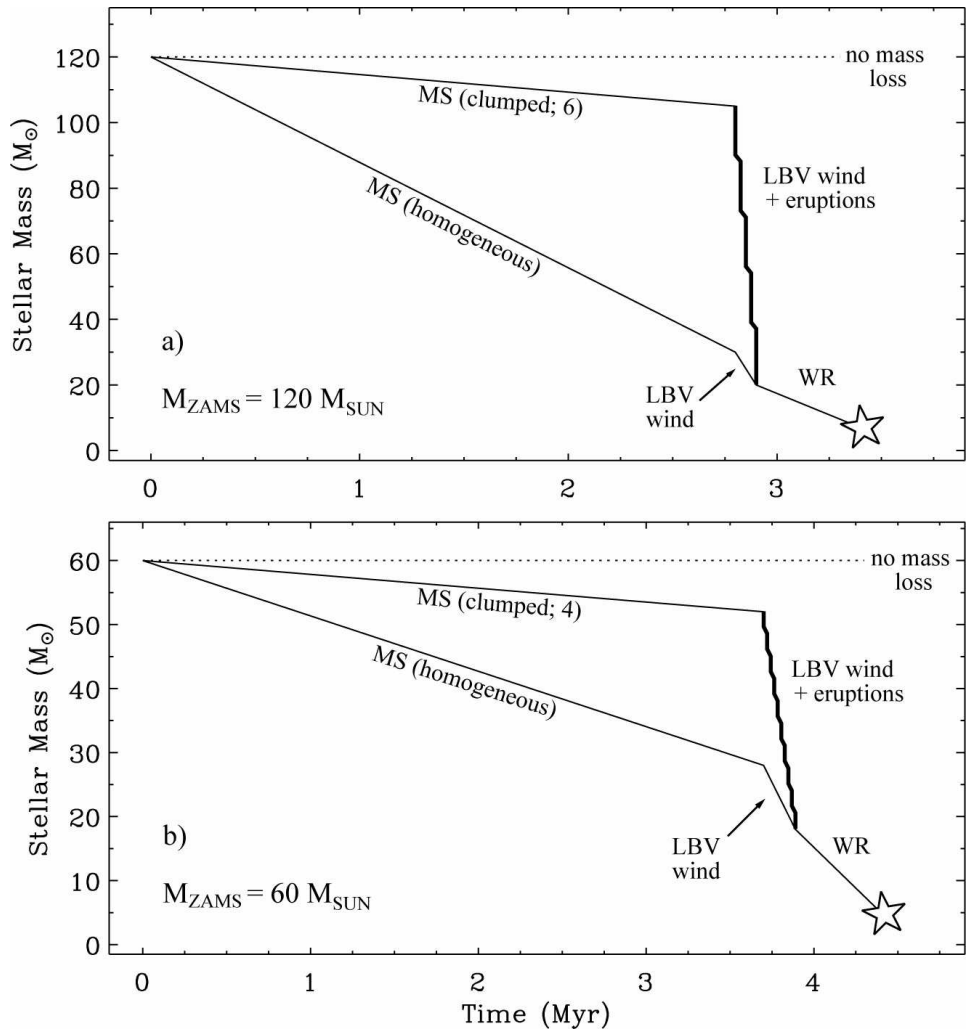,width=5.0in}
\caption{These plots are schematic representations of a star's mass as
  a function of time.  Two extreme scenarios are shown: One has higher
  conventional O-star mass-loss rates assuming homogeneous winds on
  the main-sequence (MS) with no clumping.  This is followed by a
  brief LBV wind phase and a longer WR wind phase before finally
  exploding as a supernova; this is the type of scenario usually
  adopted in stellar evolution calculations.  The second has much
  reduced mass-loss rates on the main sequence (assuming clumping
  factors of 4--6), followed by an LBV phase that includes severe mass
  loss in several brief eruptions plus a steady wind in the time
  between them; this is the type of scenario discussed by Smith \&
  Owocki (2006).  Panel (a) shows the case for an initial stellar mass
  of 120 M$_{\odot}$ (appropriate for a luminous LBV like AG Carinae),
  and Panel (b) shows an initial mass of 60 M$_{\odot}$ (appropriate
  for a somewhat less luminous object like P Cygni, perhaps).  The
  more numerous, more frequent, and less extreme eruptions of the
  60~M$_{\odot}$ scenario assume that the mass lost in an LBV burst
  may scale with proximity to the Eddington limit.  Note that the
  clumping factors of 4--6 shown here are still fairly modest compared
  to some estimates of $>$10 for O-star winds.}
\end{center}
\end{figure}

\section{The Problem: Line-Driven Winds Provide Insufficient Mass Loss}

In order to shed a massive star's envelope and reach the WR stage,
models must prescribe semiempirical mass-loss rates, which can be
scaled by a star's metallicity (e.g., Chiosi \& Maeder 1986; Maeder \&
Meynet 1994; Meynet et al.\ 1994; Langer et al.\ 1994; Langer 1997;
Heger et al.\ 2003).  Often-adopted ``standard'' mass-loss rates are
given by de Jager et al.\ (1988), and Nieuwenhuijzen \& de Jager
(1990).  In order for stellar evolution models to match observed
properties at the end of H burning, such as WR masses and
luminosities, and the relative numbers of WR and OB stars, these
mass-loss rates need to be enhanced by factors of $\sim$2 (Maeder \&
Meynet 1994; Meynet et al.\ 1994).

However, such enhanced mass-loss rates contradict observations.
Recent studies suggest that mass-loss rates are in fact 3--10 or more
times {\it lower} than the ``standard'' mass-loss rates, not higher.
This is due to the influence of clumping in the winds (Fullerton et
al.\ 2006; Bouret et al.\ 2005; Puls et al.\ 2006; Crowther et al.\
2002; Hillier et al.\ 2003; Massa et al.\ 2003; Evans et al.\ 2004;
Kudritzki \& Urbaneja in these proceedings), such that mass-loss rates
based on density-squared diagnostics like H$\alpha$ and free-free
radio continuum emission have led to overestimates if the wind is
strongly clumped.  The consequent reduced mass-loss rates mean that
steady winds are simply inadequate for the envelope shedding needed to
form a WR star.  This is not such a problem for stars below 10$^{5.8}$
L$_{\odot}$, where the red supergiant (RSG) wind may be sufficient.
However, above 10$^{5.8}$ L$_{\odot}$ (initial mass above 40--50
$M_{\odot}$) stars do not become RSGs (Humphreys \& Davidson 1979),
posing a severe problem if these stars depend upon line-driven winds
for mass loss.

For example, consider the fate of a star with initial mass of 120
M$_{\odot}$.  The most extreme O2 If* supergiant HD93129A has a
mass-loss rate derived assuming a homogeneous wind of roughly
2$\times$10$^{-5}$ M$_{\odot}$ yr$^{-1}$ (Repolust et al.\ 2004).  If
the true mass-loss rate is lower by a factor of 3--10 or more as
indicated by clumping in the wind, then during a $\sim$2.5 Myr MS
lifetime (Maeder \& Meynet 1994), the star will only shed about 5--20
M$_{\odot}$, leaving it with M$>$100~M$_{\odot}$, and an additional
80~M$_{\odot}$ deficit to shake off before becoming a WR star.  After
this, the stellar wind mass-loss rates are higher during post-MS
phases, but they are still insufficient to form a WR star.  They
therefore cannot make up for the lower $\dot{M}$ values on the MS.
For a typical LBV lifetime of a few 10$^4$--10$^5$ yr (Bohannan 1997)
and a typical $\dot{M}$ of $\sim$10$^{-4}$ M$_{\odot}$ yr$^{-1}$ for
most LBVs, the LBV phase will only shed a few additional solar masses
through its line-driven wind.  Thus, some mechanism other than just a
steady wind is needed to reduce the star's total mass by several dozen
M$_{\odot}$.

\begin{figure}\begin{center}
\epsfig{file=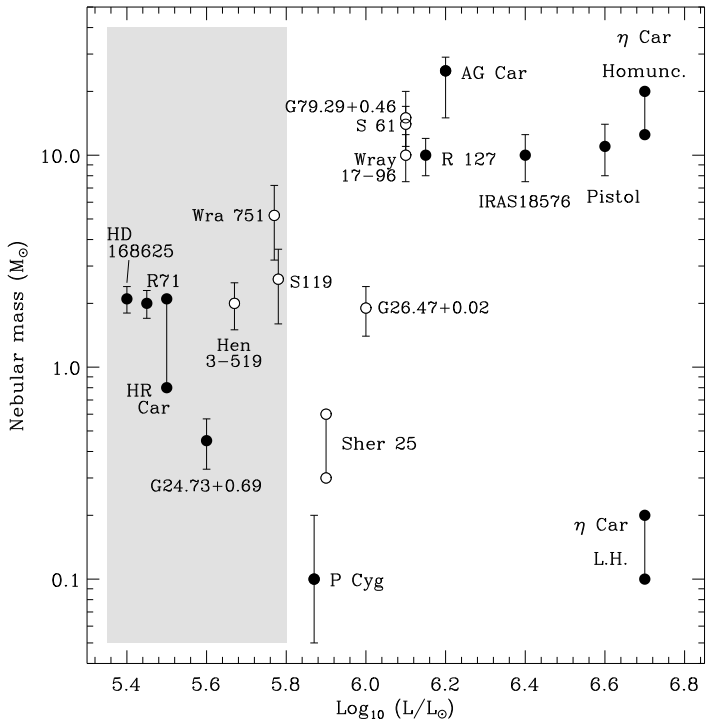,width=4.0in}
\caption{Masses of ejecta nebulae around LBVs (filled dots) and LBV
  candidates (unfilled) as a function of the central star's bolometric
  luminosity.  Luminosities are taken from Smith, Vink, \& de Koter
  (2004), while sources for the masses are given in Smith \& Owocki
  (2006).}
\end{center}
\end{figure}

\section{Balancing the Budget: LBV Eruptions}

The most likely mechanism to rectify this hefty mass deficit is giant
eruptions of LBVs (e.g., Davidson 1989; Humphreys \& Davidson 1994;
Humphreys, Davidson, \& Smith 1999; Smith \& Owocki 2006), where the
mass-loss rate and bolometric luminosity of the star increase
substantially.  While we do not yet fully understand what causes these
giant LBV outbursts, we know empirically that they do indeed occur,
and that they drive substantial mass off the star.  Deduced masses of
LBV and LBV-candidate nebulae from the literature are plotted in
Figure 2 as a function of the central star's luminosity.  We see that
for stars with log(L/L$_{\odot}$)$>$6, nebular masses of 10
M$_{\odot}$ are quite reasonable, {\it perhaps suggesting that this is
a typical mass ejected in a giant LBV eruption}.

If such large masses are typical for LBV outbursts, then only a few
such eruptions occurring sequentially during the LBV phase are needed
to remove a large fraction of the star's total mass.  This is shown
schematically in Figure 1 for stars with initial masses of 120 and 60
M$_{\odot}$.  Notice that in the 60 M$_{\odot}$ example, the LBV
eruptions are more numerous and each one is less massive than in the
120 M$_{\odot}$ case; this is entirely hypothetical, but is based on
the presumption that a more massive and more luminous star will have
more violent mass ejections because of its closer proximity to the
Eddington limit.  For example, we might expect that $\eta$ Car
currently has an Eddington parameter of $\Gamma$=0.9 or higher,
whereas a less luminous LBV like P Cygni probably has $\Gamma$=0.5 or
so.  Further investigation of the amount of mass ejected in each
burst, and their frequency and total number is probably the most
important observational pursuit associated with LBVs and their role in
stellar evolution.
 
Our best example of this phenomenon is the 19th century ``Great
Eruption'' of $\eta$ Carinae.  The event was observed visually, the
mass of the resulting nebula has been measured (12--20 M$_{\odot}$ or
more; Smith et al.\ 2003), and proper motion measurements of the
expanding nebula indicate that it was ejected in the 19th century
event (e.g., Morse et al.\ 2001).  The other example for which this is
true is the 1600 {\sc ad} eruption of P Cygni, although its shell
nebula has a much lower mass (Smith \& Hartigan 2006).  Both
$\eta$~Car and P~Cyg are surrounded by multiple, nested shells
indicating previous outbursts (e.g., Walborn 1976; Meaburn 2001).
While the shell of P~Cyg is less massive than $\eta$~Car's nebula, it
is still evident that P~Cyg shed more mass in such bursts than via its
stellar wind in the time between them (Smith \& Hartigan 2006).  This
difference between P~Cyg and $\eta$ Car hints that LBV outbursts do
indeed become progressively more extreme near the Eddington limit.
However, the Homunculus and the Little Homunculus around $\eta$~Car
also caution that any one star can eject very different amounts of
mass in each of its subsequent eruptions, with a corresponding wide
range of luminous and kinetic energy.

Although LBV eruptions are rare, a number of extragalactic $\eta$ Car
analogs or ``supernova impostors'' have been observed, such as SN1954J
in NGC2403 and SN1961V in NGC1058 (Humphreys et al.\ 1999; Smith et
al.\ 2001; Van Dyk et al.\ 2002, 2005), V1 in NGC2363 (Drissen et al.\
1997), and several recent events seen as type IIn supernovae, like
SN1997bs, SN2000ch, SN2002kg, and SN2003gm (Van Dyk et al.\ 2000,
2006; Wagner et al.\ 2004; Maund et al.\ 2006).  Furthermore, massive
circumstellar shells have also been inferred to exist around
supernovae and gamma-ray bursters (GRBs).  Some examples are the
radio-bright SN1988Z with a nebula as massive as 15 M$_{\odot}$
(Aretxaga et al.\ 1999; Van Dyk et al.\ 1993; Chugai \& Danziger
1994), as well as similar dense shells around SN2001em (Chugai \&
Chevalier 2006), SN1994W (Chugai et al.\ 2004), SN1998S (Gerardy et
al.\ 2002), SN2001ig (Ryder et al.\ 2004), GRB021004 (Mirabal et
al. 2003), and GRB050505 (Berger et al.\ 2005).

These outbursts and the existence of massive circumstellar nebulae
indicate that the 19th century eruption of $\eta$ Car is not an
isolated, freakish event, but instead may represent a common rite of
passage in the late evolution of the most massive stars. A massive
ejection event may even initiate the LBV phase, by lowering the star's
mass, raising its L/M ratio, and drawing it closer to instability
associated with an opacity-modified Eddington limit (Appenzeller 1986;
Davidson 1989; Lamers \& Fitzpatrick 1988; Humphreys \& Davidson
1994).  Mass loss in these giant eruptions may play a role in massive
star evolution analogous to thermal pulses of asymptotic giant branch
stars.  In any case, meager mass-loss rates through stellar winds,
followed by huge bursts of mass loss in violent eruptions at the end
of core-H burning (see Fig.\ 1) may significantly alter stellar
evolution models.

\begin{figure}\begin{center}
\epsfig{file=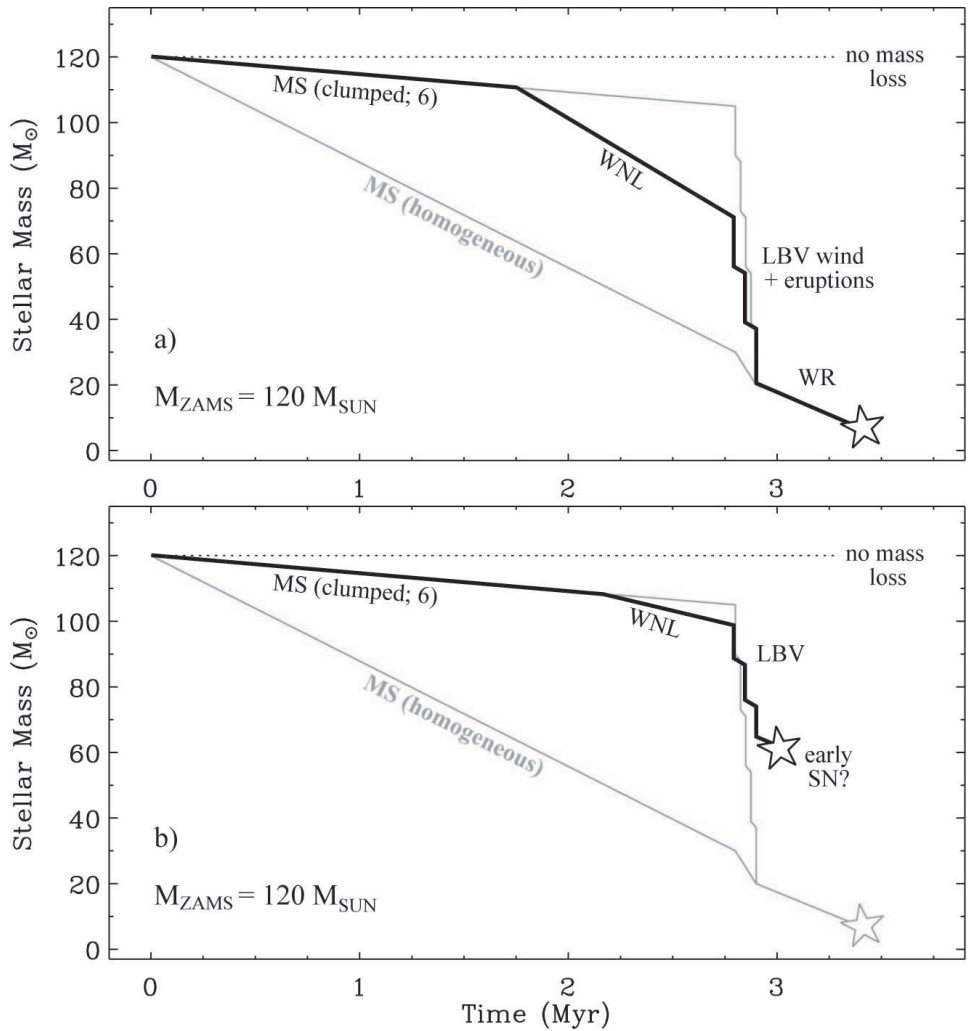,width=5.0in}
\caption{The same type of schematic plot as shown in Figure 1, but
  this figure shows some more complicated hypothetical alternatives in
  between the two extremes of Figure 1$a$ (only one case of M$_{\rm
  ZAMS}$=120 M$_{\odot}$is shown here).  The first in Panel (a) allows
  for substantial mass loss via a wind during an extended late-type WN
  phase that lasts for almost half the MS lifetime.  The total mass
  lost by the WNL wind is almost as much as through LBV eruptions, but
  one could easily adjust these depending on the duration of the WNL
  phase.  The second alternative in Panel (b) has a weaker WNL phase
  and relaxes the assumption that bonafide WR stars are the
  decsendants of the most massive stars -- i.e. it allows for the
  possibility that the very most massive stars might explode before
  reaching the WR phase, thereby reducing the amount of mass that
  needs to be shed through LBV eruptions.}
\end{center}
\end{figure}

\section{Alternative Scenarios}

The scenario where LBV eruptions dominate the mass loss of the most
massive stars, as shown in Figure 1, would represent a dramatic change
in our understanding of mass loss in stellar evolution.  Therefore, it
certainly deserves close scrutiny, and it is worth considering some
possible alternatives or modifications.  

First, however, I would like to emphasize that {\it the scenario in
which homogeneous line-driven winds of O stars dominate the mass lost
during the life of a star is almost certainly wrong}.  There are two
independent reasons to think so.

This need for recognizing the role of LBV eruptions in mass loss is
partly motivated by recent studies of the mass-loss rates of O stars,
where clumping in the winds suggests rather drastic reductions in the
MS mass-loss rates.  To be fair, the required amount of reduction in
mass-loss rates is not yet settled; some indications favor reduction
of more than an order of magnitude, while other estimates are more
moderate, indicating factors of only a few.  While this is debated, it
is worth remembering that even if the mass-loss rate reduction is only
a factor of 3, {\it it would still indicate that LBV eruptions may
dominate the total mass lost during the lifetime of a very massive
star}.  Note that the plots in Figure 2 adopt fairly modest mass-loss
rate reduction factors of only 4--6.

However, clumping in O-star winds is only part of the story.  The
other part is the observational reality that LBV eruptions like $\eta$
Car's massive 19th century outburst do indeed occur, and we have
evidence that they occur more than once, ejecting a mass of order 10
M$_{\odot}$ each time.  A star's mass budget needs to allow for that.
However, if we require several 10's of solar masses in LBV eruptions,
plus enhanced mass loss during a WNL phase (see below), we run into a
{\it serious} problem --- homogeneous winds simply do not allow enough
room for additional mass loss through WNL phases and LBV eruptions!
Thus, the mass-loss rates implied by the assumption of homogeneous
winds are not viable. I would then suggest that the existence of WNL
and LBV mass loss is an independent argument that O star winds {\it
must} be clumped, reducing their mass-loss rates by at least a factor
of 2--3.

Nevertheless, the provocative new scenario in Figure 1 may still seem
a bit extreme, placing a huge and possibly unrealistic burden on LBV
eruptions.  The truth may lie somewhere in between, so let's consider
two likely alternatives.

\subsection{A Long WNL Phase?}

One alternative is that a very massive star spends a good fraction of
its H-burning MS lifetime as a late-type WN star (WNL; see Crowther et
al.\ 1995).  Even if their winds are clumped, WNL stars have much
higher mass-loss rates than their O star counterparts.  Thus, if it is
possible that massive stars spend something like a third or half of
their MS lifetime as a WNL star, they can take a substantial chunk out
of the star's total mass.  This could temper the burden placed upon
LBVs.  This scenario is sketched in Figure 3$a$.  While Figure 3$a$
seems reasonable (even likely) to me, there are a few caveats to keep
in mind.

First, Figure 3$a$ with its rather long WNL phase could only apply to
the very most freakishly massive stars, with initial masses above
roughly 90--100 M$_{\odot}$.  The justification for this comment is
that spectral type O3 and even O2 stars still exist in clusters within
star forming regions that are 2.5--3 Myr old (like Tr16 in the Carina
Nebula).  O3 stars probably have initial masses around 80--100
M$_{\odot}$ or so, and MS lifetimes around 3 Myr.  Therefore, these
stars cannot spend a substantial fraction of their H-burning lifetime
as a WNL star, because they evidently live for about 3 Myr without yet
reaching the WNL phase.  Only for the most massive stars, which are
even more extremely rare, might a relatively long WNL phase be
possible.  This makes me wonder if we have yet another a dichotomy in
stellar evolution, with very different evolutionary sequences above
and below 100--120 M$_{\odot}$ -- much like the dichotomy above and
below 45--50 M$_{\odot}$.  One could certainly make the case that the
most luminous evolved stars that are sometimes called LBVs or LBV
candidates -- stars like $\eta$~Car, the Pistol star, HD5980, and
possibly LBV~1806-20 -- have followed a different path than the
``normal'' LBVs like AG~Car and R127.

\begin{figure}\begin{center}
\epsfig{file=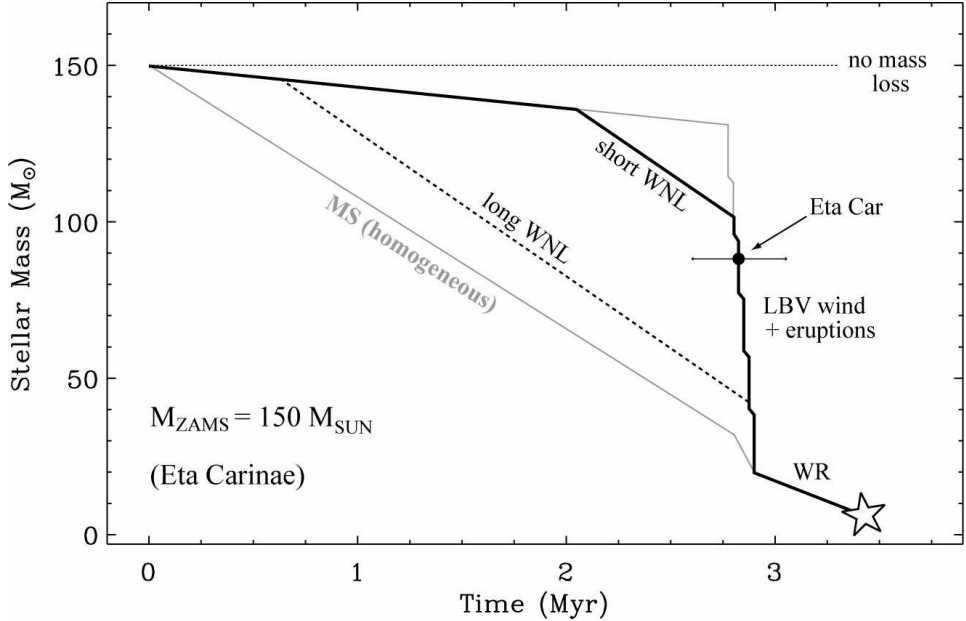,width=5.0in}
\caption{Same as Figures 1 and 3, but for a star with an initial mass
  at the upper mass limit of 150 M$_{\odot}$, perhaps appropriate for
  $\eta$ Carinae.  Here I show what the mass evolution might look like
  for a relatively short (small contribution) and a relatively long
  (dominant contribution) WNL phase, as well as the simpler extremes
  with a strong MS wind, as well as LBV eruptions with no WNL phase in
  gray.  The dot shows the likely currently-observed locus of $\eta$
  Carinae (note that I am being quite generous here with the
  correction for $\eta$ Car's companion star).  Considering that we
  know $\eta$ Car has already suffered 2--3 major LBV eruptions, which
  scenario is most consistent with its present mass?}
\end{center}
\end{figure}

Second, I would suggest that while WNL stars may make some
contribution to the mass loss at the highest luminosities, their
influence must be limited.  They cannot provide the majority of mass
lost by these stars, so the LBV eruption mass loss must still
dominate.  The reasoning behind this comment has to do with the
available mass budget of $\eta$~Carinae; namely, that $\eta$ Car is
probably a post-WNL star, {\it but it still has retained most of its
original mass}.

Let's remember that $\eta$ Car is the most luminous and most evolved
member of a rich region containing over 65 O-type stars, as well as 3
WNL stars (see Smith 2006).  It is fair to assume that the current LBV
phase of $\eta$ Car is not only a post-MS phase, but probably also a
post-WNL phase, since its ejecta are more nitrogen rich than the WNL
stars in Carina.  It is also safe to assume that $\eta$ Car has
advanced further in its evolution sooner than the WNL stars of the
same age in this region simply because it is more luminous and started
with a higher initial mass.  Now, $\eta$ Car is seen today surviving
as a very massive star of around 100 M$_{\odot}$ or more, and we
measure a total of something like 20-35 M$_{\odot}$ in its
circumstellar material ejected in only the last few thousand years
(the Homunculus, plus more extended outer material; see Smith et al.\
2003, 2005).  That means $\eta$ Car began its LBV phase -- and ended
its MS and/or WNL phase -- with more than 120 M$_{\odot}$ still bound
to the star!\footnote{Parameters could be chosen selectively to push
this as low as perhaps 100--105 M$_{\odot}$, but not lower).}  If
there really is an upper limit of about 150 M$_{\odot}$ to the mass of
stars, then {\it this rules out the possibility that winds during the
MS or WNL phases could dominate the mass-lost by the star in its
lifetime}.  Consequently, it also requires that the MS and WNL winds
were indeed highly clumped.  If folks don't like relying on just
$\eta$ Car because it is an abomination, there's the Pistol star,
which is also a post-MS object and has a present-day mass that
probably exceeds 100 M$_{\odot}$.

This argument is made graphically in Figure 4, where options of
``long'' and ``short'' WNL phases are shown.  Keeping three facts in
mind --- 1) that we see more than 20 M$_{\odot}$ of nebular material
from recent LBV eruptions around $\eta$ Car, 2) that $\eta$ Car has a
present day mass around 100 M$_{\odot}$ if it is not violating the
classical Eddington limit (I am being generous with the companion
star's mass in Figure 4), and 3) that there is a likely upper mass
limit for stars of around 150 M$_{\odot}$ --- where would you place
$\eta$ Carinae on each track in Figure 4?  What does that signify for
the relative importance of the WNL phase?

\subsection{An Early Death at the End of the LBV Phase?}

The main motivation for such huge amounts of mass loss in
continuum-driven LBV eruptions is the assumption that even the most
massive stars eventually reach the WR phase, requiring that their mass
be reduced down to about 20 M$_{\odot}$ before that point (see Smith
\& Owocki 2006).  If we can relax this constraint and say that the
most massive stars above 100 M$_{\odot}$ perhaps {\it do not} make it
to the WR phase, then we can alleviate the burden of removing so much
mass through LBV explosions.  This would be saying that the most
massive stars might undergo core collapse at the end of the LBV phase,
instead of entering the WR phase (Figure 3$b$).

That is easy to say and it would seem to fix the uncomfortable problem
of depending on LBVs for such drastic mass shedding.  However, we
should be mindful that this alternative would require an {\it even
more} radical paradigm shift in our understanding of stellar evolution
than Figure 1$a$.  Namely, Figure 3$b$ would require that not only are
LBVs in a core He burning phase\footnote{In fact, they must have
reached it before the LBV phase, because the LBV phase is so short.},
but that LBVs even reach advanced stages like core O and Si burning.
Current understanding implies that LBVs have not yet reached He
burning.  On the other hand, perhaps an early supernova explosion
during the LBV phase is not crazy after all, since we really don't
know what is going on deep inside the star.  In fact, there are
several reasons why an early explosion like in Figure 3$b$ might be
attractive:

\begin{itemize}

\item As noted earlier in \S 3, several observations of supernovae
  (especially Type IIn supernovae) and GRBs reveal that they have
  dense, massive circumstellar shells close to the star.  In their
  talks at this meeting, H.-W.\ Chen and D.\ Fox noted additional
  examples of GRBs with dense ($\sim$10$^6$ cm$^{-3}$) circumstellar
  shells seen in absorption spectra of afterglows.  In some cases
  these closely resemble the absorption features in the shell around
  $\eta$ Car (T.\ Gull, these proceedings).  Where did these compact
  and dense circumstellar shells come from if the WR phase has a
  sustained fast wind for a few 10$^5$ years?  The answer may be that
  these shells did in fact originate in LBV-like outbursts that
  occured within about 1000 years of the final death of the star.
  That would be astonishing and very important if true.

\item So far, I don't know any example of a bona-fide WR star that is
  surrounded by an extremely massive (like $\sim$50 M$_{\odot}$) group
  of nested shells left over from a previous LBV phase.  Perhaps such
  objects would be rare anyway and don't last long in an observable
  phase like this, but it would be reassuring to see at least one
  example.  If the most massive stars explode at the end of the LBV
  phase, then we wouldn't necessarily expect such massive shells
  around any WR star.
 
\item Oxygen burning is unstable, and as noted by A.\ Heger in his
  talk, can lead to short pulsational bursts that may supply
  sufficient mechanical energy to power an eruption like $\eta$ Car's
  19th century event with $\sim$10$^{50}$ ergs.  The problem with this
  scenario, though, is that the duration of O burning is extremely
  short and could not account for the observational fact that $\eta$
  Car-like eruptions tend to recur on timescales of $\sim$10$^3$
  years.  These O-burning blasts could only account for a last hurrah
  right before the star's final demise...but the possibility is
  interesting anyway.


\end{itemize}

In any case, an explosion at the end of the LBV phase when the star is
still very massive would almost certainly form a black hole, and this
should happen in roughly 3 Myr.  Are there any examples of massive
black holes in massive star clusters?  If so, where are the expanding
supernova remnants from these events?  If this scenario were true, of
course, it would mean that $\eta$~Carinae and stars like it in other
galaxies may explode as hypernovae at any moment.  This would be good
for my chances of getting future observing proposals accepted, but I
assure the reader that this is not why I am mentioning the
possibility.

\subsection{Binaries...?}

In addition to these two alternatives listed above, the potential role
of close binaries -- in particular, Roche Lobe Overflow (RLOF) -- has
been a glaring omission so far.  In a wide variety of different
scenarios depending on intial conditions, close binary evolution can
modify a star's mass (see, for example, Vanbeveren et al.\ 1998).
Given the fact that most stars are binaries, this should be considered
as well, but I don't wish to get into this complex topic here.  I
would like to note, however, that like the continuum-driven mass loss
in LBV eruptions, mass loss/tansfer through RLOF will be relatively
insensitive to metallicity compared to line-driven winds.  Therefore,
some of the comments in the next section apply to binary alternatives
as well.

\section{Potential Implications for the First Stars and their Environments}

The first stars, which should have been metal free, are generally
thought to have been predominantly massive, exhibiting a flatter
initial mass function than stars at the present epoch (e.g., Bromm \&
Larson 2004; Bromm in these proceedings).  With no metals, these stars
should not have been able to launch line-driven winds, and thus, they
are expected to have suffered no mass loss during their lifetimes.
The lack of mass loss profoundly affects the star's evolution and the
type of supernova it eventually produces (Heger et al.\ 2003), as well
as the yield of chemical elements from the first supernovae and
hypernovae that seeded the early interstellar medium of galaxies.

This view rests upon the assumption that mass loss in massive stars at
the present time is dominated by line-driven winds, for which
$\dot{M}$ can be scaled smoothly with metallicity --- but this
assumption may be problematic in view of recent observational
constraints.  As discussed above, massive shells around LBVs and the
so-called ``supernova impostors'' in other galaxies indicate that
short-duration eruptions contribute substantially -- and may even
dominate -- the mass loss of very massive stars, while steady,
line-driven winds on the MS contribute little to the total mass lost
during their lifetime.  Unlike line-driven winds, the driving
mechanism for these outbursts is probably insensitive to metallicity,
as explained in more detail by Smith \& Owocki (2006).

Since the trigger of LBV eruptions is still unidentified, one of
course cannot yet claim confidently that these eruptions will in fact
occur in the first stars.  However, the possibility that they offer a
way for low-metallicity stars to shed large amounts of mass compells
us to consider their potential influence for stellar evolution of
the first stars and their surroundings.  Some potential consequences
are listed here:

\begin{itemize}

\item If the first stars were able to shed large amounts mass through
  continuum-driven blasts at the end of MS evolution, then it could
  affect the type of explosion and the type of remnant the star leaves
  behind.  Thus, the expected relative numbers of pair instability
  explosions compared to supernovae that produce black holes or
  neutron stars as their remnants will change.  Very massive stars
  that lose enough mass may fall below the threshold for
  pair-instability supernovae, allowing much of their core metals to
  remain trapped inside a black hole or neutron star.  This change, in
  turn, will seriously alter expectations for the chemical yield
  returned to the ISM by early supernovae, and would affect the
  initial mass function (IMF) inferred from studies of very metal poor
  stars (e.g., J.\ Tumlinson, these proceedings).

\item The early ISM of galaxies was very different than it is today.
  In the early universe, elements heavier than He recycled back into
  the ISM all came from massive stars.  This is partly because of the
  IMF skewed to higher masses, but mostly because at early times,
  intermediate mass stars had not yet evolved off the main sequence to
  return C and O to the ISM as an AGB star.  If the first stars were
  able to shed large amounts mass {\it before} exploding as
  supernovae, the ISM would have been profoundly affected.  Namely,
  the pollution of the early ISM could have a substantial contribution
  of nitrogen-rich CNO ashes, since these massive stars were likely
  mixed and self-enriched due to rotation.  In other words, the early
  ISM may have been similar to the N-rich material in circumstellar
  LBV shells seen today. This could significantly affect the dust
  content of the early ISM as well, especially for the generation of
  stars immediately following the first stars.

\item Continuum-driven blasts at the end of MS evolution might enable
  the first stars to reach and pass through a WR phase.  In that case,
  the self-enrichment of CNO products in WR atmospheres would likely
  allow them to have line-driven winds, providing even further mass
  loss (Vink \& de Koter 2005; Eldridge \& Vink 2006). In addition to
  giving us further complications in determining the end product of
  stellar evolution for Population III stars (pair instability, BH,
  NS), the existence of a WR phase in the first stars would affect the
  mechanical energy of the surrounding ISM and would contribute
  additional N and C, not to mention affecting the immediate
  circumstellar environment into which a GRB shock expands.

\end{itemize}

In short, if mass loss of massive stars at the present epoch is
dominated by mechanisms that are insensitive to metallicity, then we
must question the prevalent notion that the first stars did not lose
substantial mass prior to their final supernova event.  If these
outbursts can occur at low metallicity, it would profoundly alter our
understanding of the evolution of the first stars and their role in
early galaxies.

\acknowledgments

I thank Stan Owocki for many relevant discussions, and I thank Paul
Crowther and Peter Conti for repeatedly reminding me of the potential
importance of WNL stars.  I was supported by NASA through grant
HF-01166.01A from STScI, which is operated by the Association of
Universities for Research in Astronomy, Inc., under NASA contract
NAS5-26555.


\end{document}